# Unraveling the spin polarization of the $\nu = 5/2$ fractional quantum Hall state

L. Tiemann[1,2*], G. Gamez[1], N. Kumada[1], and K. Muraki[1,2*]

[1] NTT Basic Research Laboratories, NTT Corporation, 3-1 Morinosato-Wakamiya, Atsugi 243-0198, Japan

[2] ERATO Nuclear Spin Electronics Project, Japan Science and Technology Agency (JST)

The fractional quantum Hall (FQH) effect at filling factor $\nu = 5/2$ has recently come under close scrutiny, as it may possess quasi-particle excitations obeying nonabelian statistics, a property sought for topologically protected quantum operations. Yet, its microscopic origin remains unidentified, and candidate model wave functions include those with undesirable abelian statistics. Here we report direct measurements of the electron spin polarization of the $\nu = 5/2$ FQH state using resistively detected nuclear magnetic resonance (NMR). We find the system to be fully polarized, which unambiguously rules out the most-likely abelian contender and thus lends strong support for the $\nu = 5/2$ state being nonabelian. Our measurements reveal an intrinsically different nature of interaction in the first-excited Landau level underlying the physics at $\nu = 5/2$.

[*] To whom correspondence should be addressed. E-mail: lars.tiemann@gmail.com, muraki.koji@lab.ntt.co.jp

In a two-dimensional electron system (2DES) placed under a strong perpendicular magnetic field $B$, the interplay between quantum mechanics and the inter-electron interaction produces spectacular effects at low temperature. When the filling factor $\nu$ (= $n_s/n_\phi$), or the ratio between the numbers of electrons $n_s$ and magnetic flux quanta $n_\phi = (e/h)B$, matches magic rational values $p/q$ ($p$, $q$: integer), an energy gap forms based purely on electron correlation and makes the system's transverse (Hall) resistance invariant to small perturbations (*1*). In these fractional quantum Hall (FQH) phases, small variations of $n_s$ or $n_\phi$ generate quasi-particles with a fractional charge $e^* = \pm e/q$ (*2*). A paradigm for understanding the FQH effects is the composite-fermion model (*3*), which transforms an electron into a fictitious particle by merging it with an even number of magnetic flux quanta. The $\nu = 5/2$ FQH state (*4*), the only FQH state observed to date at even-denominator filling and incomprehensible within this simple composite-fermion picture, has recently come under close scrutiny as it may harbor something even more tantalizing: a nonabelian state of matter (*5-7*). Charge excitations from this potential nonabelian state expected at $\nu = 5/2$ would be carried by quasi-particles whose interchange takes the system from one of its many ground states to another, whereas the interchange of ordinary abelian quasi-particles only adds a phase to their wave functions. This unusual property, specific to nonabelian states of matter, is proposed as the foundation for topological quantum computation that would be robust against environmental decoherence (*8*). Despite the considerable experimental effort that has been invested in trying to find indications of its nonabelian nature (*9-13*), the results remain elusive, in that they are unable to discriminate clearly between different theoretical models, which include both those with nonabelian statistics (*5*, *14*, *15*) and those with abelian statistics (*16*).

The recently demonstrated quasiparticle charge of $e^* = e/4$ (*9, 11, 12*) is a property shared by all paired states at $\nu = 5/2$ (*17*), and it thus does not allow us to screen the $\nu = 5/2$ state for its statistics or the (model) wave function by which it is described. We address this issue by measuring the electron spin polarization, an attribute (*18-20*) that can allow us to discriminate clearly between the two relevant and most probable candidates that emerged through quasi-particle tunneling experiments (*10*): the nonabelian anti-Pfaffian state (*14, 15*), which is fully polarized, and the abelian (331) state (*16*), which is unpolarized (*19, 20*). We find a fully polarized $\nu = 5/2$ FQH state over a wide range of electron densities, thus unambiguously excluding the abelian candidate, and lending strong support for the theoretical prediction of a nonabelian state of matter at $\nu = 5/2$. Moreover, our measurements reveal spin polarization behavior in the first-excited orbital level (i.e., $N = 1$ Landau level), where the $\nu = 5/2$ FQH state emerges, that is intrinsically different from that in the lowest ($N = 0$) Landau level, where conventional FQH states reside, providing new insight into the interaction physics underlying this unique many-body state.

Our sample is a 100-μm-wide Hall bar (Fig. 1A) with the 2DES confined to a 27-nm-wide gallium arsenide (GaAs) quantum well. A heavily doped GaAs layer acts as a back gate and enables us to tune the electron density $n$ over a wide range between 0.5 and $4.2 \times 10^{15}$ m$^{-2}$. At $n = 4.2 \times 10^{15}$ m$^{-2}$, the longitudinal resistance $R_{xx}$ and the Hall resistance

$R_{xy}$ show a well-developed ν = 5/2 FQH state at 12 mK, along with pronounced FQH features at ν = 7/3 and 8/3, indicating a high sample quality (*21*) (Fig. 1B).

Our measurement of the spin polarization exploits the (hyperfine) interaction that intrinsically exists between the magnetic moments of the atoms constituting the GaAs quantum well and the spins of the electrons confined therein. When the 2DES has a non-zero spin polarization *P*, nuclei in contact with the 2DES experience a local magnetic field, which shifts their nuclear resonance frequency to a lower value by an amount proportional to *P* (Knight shift $K_s$). We measure $K_s$ using the resistively detected nuclear magnetic resonance (RD-NMR) technique (*22*). In RD-NMR, the resonant absorption of rf excitations and the resultant change in the nuclear polarization $<I_z>$ are detected as a change in $R_{xx}$ that originates from the coupling between $<I_z>$ and the electron Zeeman energy $E_Z \propto (B + b_0 \cdot <I_z>)$ mediated by the hyperfine interaction ($b_0$: constant). In order to deduce *P* from the measured NMR spectra, we need the bare unshifted resonance frequency of the nuclei and a proportionality factor between *P* and $K_s$. Therefore, in addition to ν = 5/2, we also measure states with known spin polarizations at ν = 2 and 5/3 at the same magnetic field, which respectively serve as references for zero and maximum polarization. The standard RD-NMR technique, however, is applicable to neither of these states; at ν = 2 and 5/3 $R_{xx}$ vanishes and at ν = 5/2 it does not show any response other than non-resonant heating. For that reason, we use the back gate to switch to a different filling factor $ν_{read}$ with finite $R_{xx}$ and sensitivity to readout the resonant rf absorption at ν as a change in $R_{xx}$ at $ν_{read}$ (Fig. 1C) (*23*). We examined various ν and found $ν_{read}$ = 0.59-0.65 to be most sensitive. At this particular range of filling factors near ν = 2/3 and 3/5, competition between FQH ground states with different spin polarizations makes the system behave like an Ising ferromagnet. Consequently, $R_{xx}$ becomes highly sensitive to tiny perturbations in the Zeeman energy, making $\nu_{read}$ a sensitive detector of $\langle I_z \rangle$ (*24*) (See Supporting Online Material for more details about this method). We have verified that the $\nu = 5/2$ FQH state remains well developed during our NMR measurements (details in Fig. 1D).

Figure 1D shows resonance spectra of the abundant $^{75}$As nuclei measured for three different filling factors at $B = 6.4$ T. The top spectrum was measured at $\nu = 2$, where the two lowest Landau levels ($N = 0, \uparrow$) and ($N = 0, \downarrow$) are fully occupied (right inset). At $\nu = 2$, the 2DES is unpolarized as the number of spin-up electrons matches the number of spin-down electrons. The peak's center thus marks the bare (unshifted) resonance frequency of the $^{75}$As nuclei. When the RD-NMR measurement is performed at $\nu = 5/2$ (= 2 + 1/2), we find that the resulting resonance spectrum is Knight shifted to a lower frequency, indicating a non-zero spin polarization. To determine the value of this polarization, we compare the $\nu = 5/2$ spectrum with the resonance spectrum of a state with a *known* spin polarization. The $\nu = 5/3$ (= 2 − 1/3) FQH state can be used for this purpose, as its effective spin polarization is that of the $\nu = 1/3$ Laughlin state (*2, 25*), which is always maximally polarized. Note that we accessed different filling factors by changing the number of electrons contributing to the spin polarization while keeping the magnetic field constant, as shown schematically in the right inset of Fig. 1D. Thus, the Knight shift for $\nu = 5/2$ should be 1.5 times larger than that for $\nu = 5/3$, if both states are fully spin polarized, and this is what we observed.

For an accurate determination of the spin polarization, we need to take account of the spectral shape that differs for each $\nu$. The spectrum at $\nu = 2$ is symmetric and narrow because all nuclei have the same resonance frequency without a Knight shift. The asymmetrically broadened spectral shape observed at $\nu = 5/2$ and $5/3$, on the other hand, reflects the local electron density (determined by the probability density of the subband wave function $|\Psi_\nu(z)|^2$) that varies spatially along the $z$ direction, shifting the resonance frequency of individual nuclei by a different amount. In order to fit these spectra, we therefore calculate the Knight shift $K_s(z) = \alpha_\nu \cdot |\Psi_\nu(z)|^2$ for each nucleus and then integrate over the contributions of all nuclei in the quantum well with $\alpha_\nu$ the fitting parameter representing the size of the Knight shift (see Supporting Online Material for details). The spin polarization $P$ can be calculated from the ratio $\alpha_{5/2}/\alpha_{5/3}$ while taking into account the electron densities contributing to the polarization. In agreement with our previous estimate, we find $\alpha_{5/2}/\alpha_{5/3} = 1.56$, corresponding to a polarization ratio of $P_{5/2}/P_{5/3} = 1.04$. This clearly shows that the 2DES is fully spin polarized when the $\nu = 5/2$ FQH state is formed at a sufficiently low temperature.

Although our NMR measurements have demonstrated a fully spin polarized $\nu = 5/2$ FQH state at 6.4 T, an open question is whether it remains fully polarized as the Zeeman energy is decreased. When we fit the $\nu = 5/2$ spectra measured at different magnetic fields between 4.0 and 7.0 T, the obtained parameter $\alpha$ follows a linear function reflecting the density of electrons contributing to the Knight shift that is proportional to $B$ (Fig. 2A). This indicates that the spin polarization is constant, independent of the Zeeman energy between 4.0 and 7.0 T (Fig. 2B). Mapping $R_{xx}$ over the same range of magnetic fields shows that the ν = 5/2 FQH minimum evolves continuously between 3.4 and 7.0 T (Fig. 2C). The absence of a singularity is consistent with the system being polarized over the entire density range.

With these measurements, which provide the first direct evidence of full spin polarization at ν = 5/2, we are able to exclude all unpolarized or partially polarized states as the correct description of the ν = 5/2 state. Most importantly, we are able to unambiguously rule out the abelian (331) state (*16*), which had remained one of the most likely candidates through quasi-particle tunneling experiments (*10*). Other abelian models had already been ruled out (*10*). Therefore, the remaining most likely models contain only nonabelian states, the anti-Pfaffian state (*14*, *15*), its particle-hole conjugate the Pfaffian state (*5*), and the $U(1) \times SU_2(2)$ state (*26*), which are all spin polarized and therefore consistent with our results. Although our NMR measurement cannot distinguish Pfaffian and anti-Pfaffian states, which are degenerate in the bulk without Landau-level mixing, the combination of our results and experiments on the edge properties (*10, 13*) suggests the anti-Pfaffian state as the most likely description of the ν = 5/2 state.

We note that recent optical experiments (*27, 28*) indicate an unpolarized state at ν = 5/2. In an attempt to make these results consistent with the nonabelian More-Read Pfaffian model, it has been suggested (*29*) that at low magnetic fields, the intrinsically spin-polarized ν = 5/2 system may depolarize due to the presence of spin-textured quasi-particles, known as Skyrmions, which arise from competition between the Zeeman and

Coulomb energies. However, our NMR measurements did not find an unpolarized ν = 5/2 system at 4.0 T, where according to Ref. 29, the Zeeman to Coulomb energy ratio of 0.012 would allow for the formation of Skyrmions with the help of disorder. Skyrmions should also lead to a rapid nuclear spin relaxation rate $T_1^{-1}$ through their low-energy spin fluctuations (*24, 30*). However, we observe a very low $T_1^{-1}$ <= 0.0005 s$^{-1}$ at and around ν = 5/2, which renders the existence of Skyrmions unlikely. A more plausible explanation for the observed depolarization would be that these optical measurements probe the properties of excitations *above the ground state,* where photo-generated holes carrying spin entail a considerable perturbation of the 2DES and its spin polarization. While the difference between bulk and edge phenomena has been suggested as an alternative explanation (*31*), we emphasize that our NMR measurements probe the bulk property, clearly demonstrating that the ν = 5/2 ground state is spin polarized in the bulk.

To better understand the physics behind the formation of the ν = 5/2 FQH state and its polarization, we have explored other states in the *N* = 1 and *N* = 0 Landau levels. We have mapped out the entire range of experimentally accessible filling factors at 6.4 T to compare the behavior of the Knight shift (represented by parameter α) (Fig. 3A) and the spin polarizations (Fig. 3B). In the *N* = 0 Landau level between ν = 2 and 1/3, α exhibits a complicated oscillatory pattern reflecting the various composite fermion ground states, a behavior also observed in previous studies in this regime (*32*). The dashed line in Fig. 3A represents the value expected for full polarization at each ν, calculated based on three data points near ν = 5/3. In addition to ν = 5/3, other fully polarized quantum Hall states such as ν = 1, 2/5, and 1/3 show maxima grazing or approaching the line of full polarization, whereas unpolarized FQH states such as ν = 4/3 exhibit minima approaching zero. By contrast, in the $N = 1$ Landau level, α only increases linearly for ν ≥ 2.2 across ν = 7/3, 5/2 and up to 8/3, indicating that the system is fully polarized over the range of the $N = 1$ Landau level investigated here (Fig. 3B).

Since this measurement is performed in a static magnetic field, neither the Zeeman energy $E_Z$ ($\propto B$) nor its ratio to the Coulomb energy $E_C$ ($\propto B^{1/2}$) can account for these qualitative differences between the $N = 0$ and $N = 1$ Landau levels. Thus, the factor underlying the contrasting behavior should be the *character* of the interparticle interaction, described by the Haldane pseudopotential (*33*), which is different between these Landau levels. Striking examples can be found at half fillings. Composite fermions do not pair up at and near ν = 1/2 or 3/2 due to the strong repulsion at short distances (*34*), and the system is well described as a compressible Fermi sea of composite fermions, which naturally accounts for its partial polarization (*32*). At ν = 5/2, the weak repulsion at short distances is believed to be essential for the pairing of composite fermions and the formation of the ν = 5/2 FQH state at low temperature (*17, 34*).

The properties of unpaired composite fermions at ν = 5/2, the parent state of the ν = 5/2 FQH state, can be examined through RD-NMR measurements at elevated temperatures (Figs. 4A, B). As Fig. 4A shows, the ν = 5/2 FQH state has vanished at 150 mK. The spin polarization, on the other hand, is unaffected, and nearly full polarization is retained even up to 200 mK, before it starts to gradually drop (Fig.4C), whereas the polarization at ν = 3/2 remains constant. The robustness of the high polarization to an increase in temperature is seen not only at ν = 5/2 but also in the entire $N = 1$ Landau level (Fig. 4D). Fitting the data in Fig. 4C using a simple non-interacting composite fermion model (*35*) yields a composite-fermion effective mass of $(2.68 \pm 0.24)m_e$ and $(0.71 \pm 0.05)m_e$ for ν = 5/2 and 3/2, respectively (where $m_e$ is the electron mass in vacuum). In this simple picture, the high spin polarization is explained in terms of the large effective mass of composite fermions in the $N = 1$ Landau level making the Fermi level smaller than the Zeeman energy. It is thus tempting to consider the large effective mass to be an essential ingredient in the pairing of composite fermions and the formation of the ν = 5/2 FQH state at low temperature.

Our NMR experiments have demonstrated maximal spin polarization for the ν = 5/2 FQH state over a wide range of electron densities. These measurements constitute the first direct probe of its electron spin polarization and are consistent with the nonabelian Pfaffian state (*5*) and its particle-hole conjugate, the anti-Pfaffian state (*14, 15*), while unambiguously ruling out the unpolarized (331) state (*16*), which had been the most-likely abelian contender (*10*). Moreover, our measurements reveal the unique character of the interparticle interaction in the $N = 1$ Landau level that underlies the ν = 5/2 FQH physics. Our finding thus delivers another important piece of evidence that will help unravel the ν = 5/2 puzzle. The exciting prospect of topologically protected quantum operations using the ν = 5/2 FQH state is only constrained by the pending direct experimental demonstration of its nonabelian nature.

## Author Contributions

L. T. performed all the RD-NMR and transport measurements presented in this paper. G. G. performed transport measurements to optimize the samples. N. K. advised on the RD-NMR measurements. K. M. grew heterostructures and fabricated samples. L. T. and K. M. analyzed the data and wrote the paper.

**FIGURE 1 | Experimental setup and RD-NMR spectra.**

**A**, Schematic illustration of our experimental setup. A heavily doped GaAs layer 1 μm below the 2DES serves as a back gate, allowing us to tune the electron density $n$. The sample, located inside the mixing chamber of a dilution refrigerator, is surrounded by a three-turn coil that is connected to a frequency generator. **B**, Magnetotransport characteristics of the sample at 12 mK ($n = 4.2 \times 10^{15}$ m$^{-2}$ and $\mu = 1140$ m$^2$/Vs). Transport measurements were performed with a standard lock-in technique using an ac current of 4 nA and a frequency of 17 Hz. **C**, Sequence of our RD-NMR measurement (see main text and Supporting Online Material for details). **D**, Resonance spectra of the $^{75}$As nuclei for the transition between the $I_z = 1/2$ and -1/2 nuclear Zeeman levels, measured at $B = 6.4$ T and 12 mK for ν = 2, 5/2, and 5/3 (other quadrupole-split transitions are outside the frequency range shown here). The solid line for ν = 2 is a Gaussian fit, providing the bare frequency $f_0 = 46.3997$ MHz and the linewidth $\Gamma = 1.3$ kHz. The solid lines for ν = 5/2 and 5/3 are numerical fits, from which the parameter α (proportional to the Knight shift) is deduced. **Top-left inset**: $R_{xx}$ measured without rf excitation and under continuous rf excitation with -17 dBm, the output power we used to obtain spectra. The ν = 5/2 FQH state remains reasonably well developed under continuous rf excitation. **Right inset**: Illustration of the energy levels and their occupation. For ν = 5/2 and 5/3, the diagrams illustrate the case with maximal polarization. The number of electron spins that can contribute to the Knight shift (highlighted in red) is 1.5 times larger at ν = 5/2 than at ν = 5/3.

**FIGURE 2 | Magnetic-field dependence of Knight shift and spin polarization at ν = 5/2.**

**A**, Parameter α deduced from the fitting of resonance spectra measured at ν = 5/2 at different magnetic fields. The solid line represents a function $\alpha_{full}(B) \propto B$ obtained from a linear fit of the data. As the electron density contributing to the Knight shift for a given ν is proportional to *B*, the linear *B* dependence of α indicates that the spin polarization is constant. (For the estimation of error bars, see Supporting Online Material.) **B**, Spin polarization *P* calculated as $P = \alpha/\alpha_{full}(B)$. **C,** Mapping of $R_{xx}$ plotted as a function of *B* and filling factor ν.

**FIGURE 3 | NMR measurements at different filling factors.**

**A**, Knight shift represented as parameter α plotted as a function of ν, deduced from fitting the spectra measured at 6.4 T and 12 mK. (Relevant spectra are included in the Supporting Online Material). The dashed line represents the value $\alpha_{full}(\nu)$ expected for full polarization at each ν, calculated based on three data points near ν = 5/3. (See Supporting Online Materials for details of the analysis and the estimation of error bars.) **B**, Spin polarization *P* calculated from the obtained α and the above function via $P = \alpha/\alpha_{full}(\nu)$. **C**, $R_{xx}$ versus ν, measured under continuous application of rf with the same power (-17dBm) as used for the RD-NMR measurement.

**FIGURE 4 | Temperature dependence of Knight shift and spin polarization.**

**A**, Magnetotransport characteristics at different temperatures and 6.4 T. **B**, Resonance spectra measured at $\nu = 5/2$ at four representative temperatures. **C**, Parameter $\alpha$ deduced from the fitting of resonance spectra measured at $\nu = 5/2$ and $3/2$ at different temperatures. The dashed line represent fitting based on non-interacting composite-fermion model (see Ref. (*35*) and Supporting Online Material for details), which yields a composite-fermion effective mass of $(2.68 \pm 0.24)m_e$ ($\nu = 5/2$) and $(0.71 \pm 0.05)m_e$ ($\nu = 3/2$), where $m_e$ is the electron mass in vacuum. **D**, Parameter $\alpha$ for various $\nu$ in the $N = 1$ Landau level obtained at different temperatures. The solid line represents the value expected for full polarization as determined from the data in Fig. 3A.

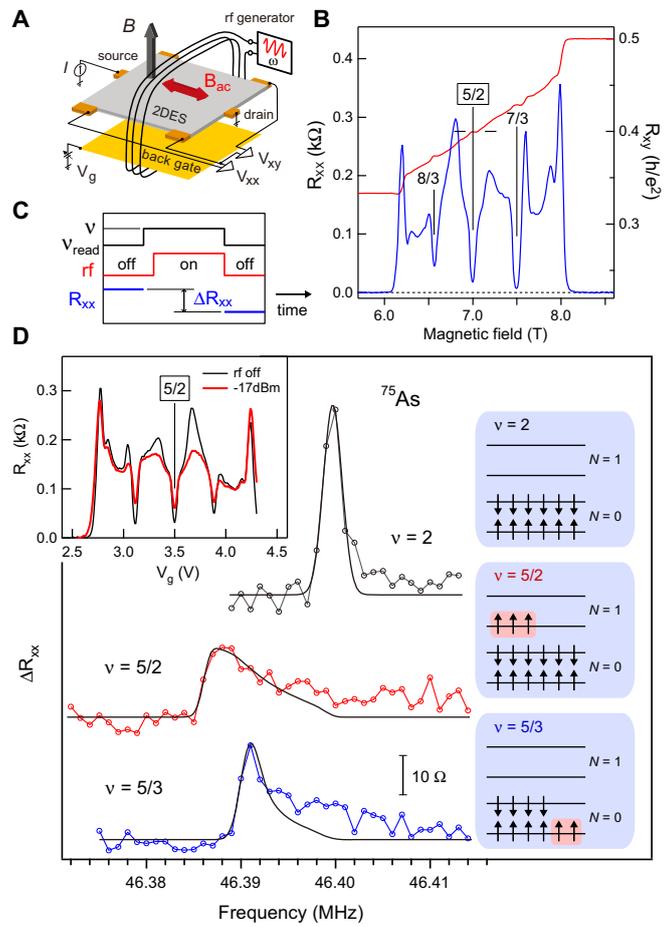

Fig. 1

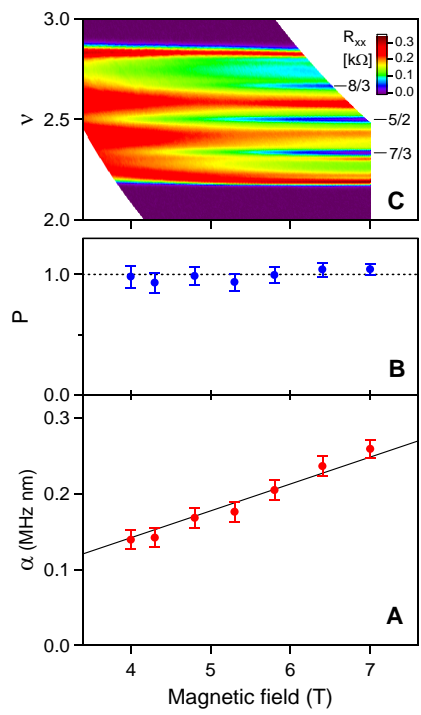

Fig. 2

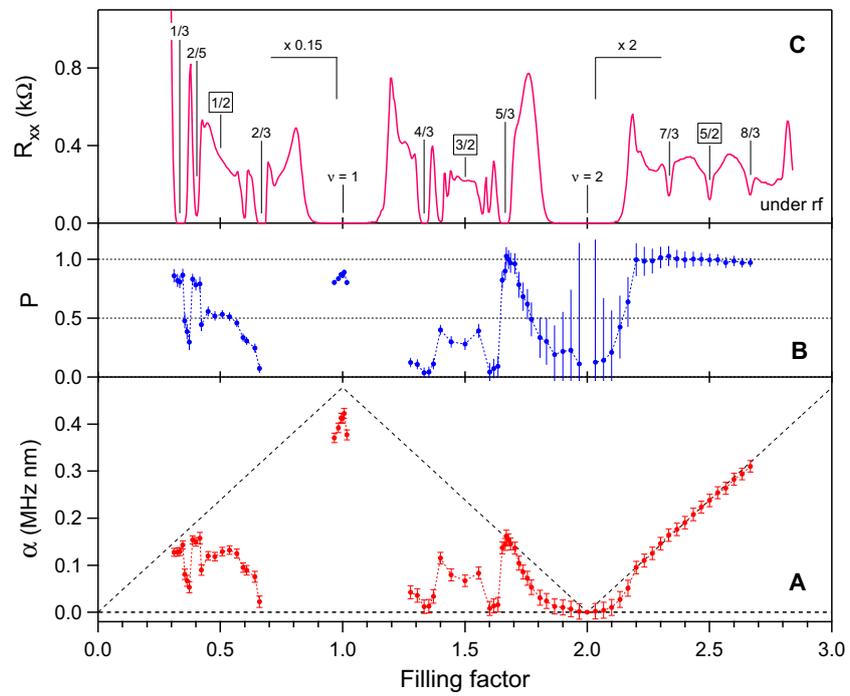

Fig. 3

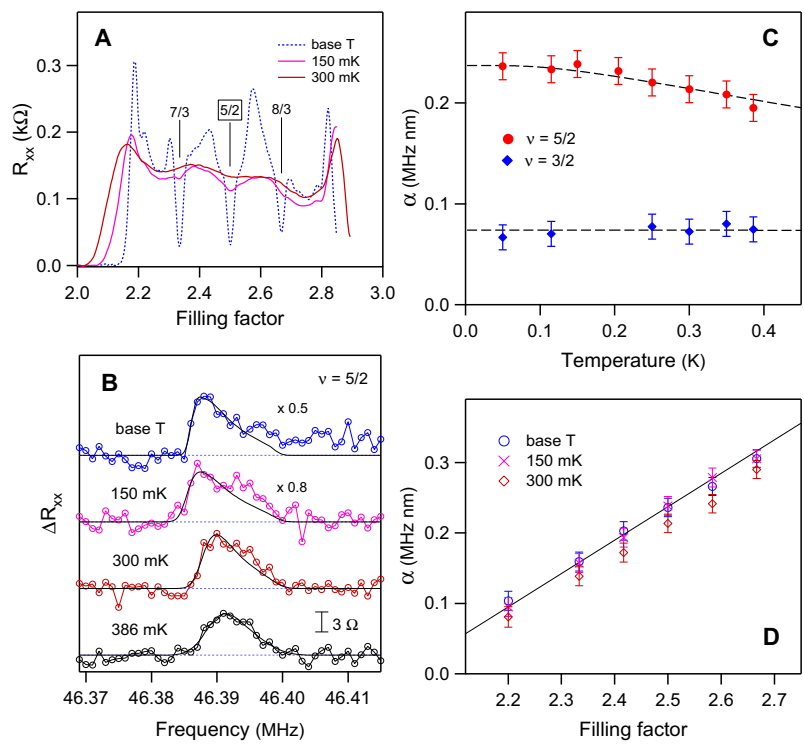

Fig. 4

## Supporting Online Material

**Sample**

The 2DES is confined to a 27-nm-wide GaAs/Al$_{0.25}$Ga$_{0.75}$As quantum well grown by molecular beam epitaxy. The sample is δ-doped on one side (front side) of the quantum well at a setback distance of 90 nm, which provides the 2DES with density $n = 1.55 \times 10^{15}$ m$^{-2}$ and mobility $\mu = 580$ m$^2$/Vs in the as-grown condition ($V_g = 0$ V). The heavily doped GaAs layer 1 μm below the 2DES serves as a back gate, which allows us to tune $n$ between 0.5 and $4.2 \times 10^{15}$ m$^{-2}$. The mobility exceeds 1000 m$^2$/Vs for $n \geq 2.8 \times 10^{15}$ m$^{-2}$, reaching a maximum value of 1150 m$^2$/Vs around $n = 3.9 \times 10^{15}$ m$^{-2}$. The well width of $w = 27$ nm was determined from cross-sectional transmission electron microscopy (TEM) on a simultaneously fabricated Hall-bar chip right next to the one used for the measurements. Details of the sample optimization and the processing are described in Refs. (*SR1*) and (*SR2*), respectively.

**RD-NMR measurements**

As illustrated in Fig. 1C, our measurement sequence consists of three steps: i) dynamical nuclear polarization at $\nu_{read}$ (4 min.), ii) rf irradiation at $\nu$, and iii) readout at $\nu_{read}$ (10-20 sec.). During periods i) and iii), a current of 45 nA is driven to dynamically polarize nuclei and to read $R_{xx}$. During period ii), the current is turned off and the coil is activated for 0.3 - 10 seconds after an initial waiting time of 110 ms. The difference ($\Delta R_{xx} = R_{xx}^{(i)} - R_{xx}^{(iii)}$) in $R_{xx}$ right before and after the period ii) is our signal. By repeating this sequence for different frequencies, we obtain a resonance spectrum.

The filling factor $\nu_{read}$ used for the readout was optimized for each magnetic field so that the best RD-NMR signal is obtained, and it ranges from $\nu_{read} = 0.59$ ($B = 6.4$-$7.0$ T) to 0.65 ($B = 4.0$ T). In addition to its high sensitivity, this range of $\nu_{read}$ has the advantage that nuclear spins can be dynamically polarized by applying a current through the sample, thus enhancing the RD-NMR signal. This dynamical nuclear polarization results from the electron-nuclear spin flip-flop scattering at boundaries between regions with differently polarized electron systems, characteristic of Ising ferromagnetic behavior in this range of $\nu_{read}$ (*SR3*). Our RD-NMR measurement is limited to temperature below 400 mK. At even higher temperatures, sufficiently large nuclear polarization to produce a RD-NMR signal could not be generated.

**Fitting of RD-NMR spectra**

The asymmetrically broadened spectral shape observed for those states with non-zero spin polarization reflects the local electron density (determined by the probability density of the subband wave function $|\psi_\nu(z)|^2$) that varies spatially along the $z$ direction, shifting the resonance frequency of individual nuclei by a different amount. Such an asymmetrically broadened spectral shape is commonly observed in NMR of 2DESs and is well documented in the literature (*SR4*). In order to obtain an accurate measure of the Knight shift that represents the measured spectra for each $\nu$, we fit these spectra by calculating the Knight shift $K_s(z) = \alpha_\nu \cdot |\psi_\nu(z)|^2$ for each nucleus and then integrate over the contributions of all nuclei in the quantum well with $\alpha_\nu$ the parameter representing the size of the Knight shift.

In RD-NMR, nuclear spin flips induced by resonant rf absorption are detected as a change in $R_{xx}$. The measured spectral shape thus reflects the local readout sensitivity of the 2DES that varies with the probability density of the 2DES at each position of the nuclei. As we use a state at a different filling factor $\nu_{read}$ (= 0.59-0.65) for the readout, its subband wave function $\psi_{read}(z)$ must therefore be taken into account as well.

We assume that the absorption spectrum of individual nuclei is described by a Gaussian function $g(f - f_{res}) = A \cdot \exp[-(f - f_{res})^2/\Gamma^2]$ with $f_{res}$ the resonance frequency and $\Gamma$ the linewidth.

The RD-NMR spectrum measured at ν is then expressed in the form of

$$I_\nu(f - f_0) = \int_{-w/2}^{w/2} g(f - f_0 + \alpha_\nu |\psi_\nu(z)|^2) |\psi_{read}(z)|^2 \, dz, \quad (1)$$

where $w$ is the quantum well width and $f_0$ the bare unshifted resonance frequency, and the factor $|\psi_{read}(z)|^2$ accounts for the position dependence of the readout sensitivity.

When fitting RD-NMR spectra taken at a given magnetic field, we first fit the spectrum at ν = 2 using a Gaussian function to find $f_0$ at that field. [Setting $\alpha_\nu = 0$ in Eq. (1) reveals that the spectrum at ν = 2 coincides with the Gaussian function assumed.] Then, the spectra for other ν are simulated using Eq. (1) with $\psi_\nu(z)$ and $\psi_{read}(z)$, which are numerically calculated based on $V_g$'s for ν and $\nu_{read}$ (and the electron density at $V_g = 0$ V) (see Fig. S1A). We obtain $\alpha_\nu$ by fitting the measured spectrum using Eq. (1) with $\alpha_\nu$, Γ and $A$ the fitting parameters. Note that $A$ is in principle independent of ν; indeed, fitting generally yields nearly identical values of $A$ for all ν. Except for particular range of ν (see Fig. S2), fitting also yields similar values of Γ, typically less than 2 kHz, for all ν including ν = 2. This indicates that the electron system is homogeneous in the two-dimensional plane and the Knight shift does not induce additional broadening other than that due to the subband wave function.

Figure S1B shows spectra at $B$ = 6.4 T calculated for various ν with $\nu_{read}$ = 0.59 and Γ = 1.5 kHz. The top panel shows the spectra for ν = 5/2 calculated using different values of α corresponding to a polarization between 0 and 1. The calculations illustrate how the spectrum becomes increasingly asymmetric and broad as the polarization increases. The middle panel compares the spectra for ν = 1/3, 5/3, and 7/3 calculated with the same parameters except $\psi_\nu(z)$. Despite the same value of α (corresponding to full polarization of these states), the calculations predict that the spectrum for ν = 1/3 shows a much larger Knight shift, in qualitative agreement with experiment (Fig. S2). This is due to $|\psi_\nu(z)|^2$ being strongly squeezed at ν = 1/3 by the large negative back gate voltage, which leads to the high peak in $|\psi_\nu(z)|^2$ and hence a large Knight shift. The spectra for ν = 1 and 8/3, shown in the bottom panel, predict characteristic features, a sharp peak and a long tail at ν = 1 and a broad spectrum without a sharp peak at ν = 8/3, both of which are observed in experiment (Fig. S2).

Among the spectra shown in Fig. S2, those at ν = 2/5 and 0.373 show significant broadening, with Γ = 2.8 and 5.3 kHz, respectively. We speculate that between ν = 1/3 and 2/5 the electron system becomes inhomogeneous in the two-dimensional plane as a result of strong disorder at such low densities. While this also affects the spectrum at ν = 2/5, the absence of such additional broadening at other ν, in turn, indicates that the electron system is homogenous at other ν shown in Fig. 2S.

**Relation between α and spin polarization $P$**

The spin polarization $P$ is calculated as the ratio between the obtained α and the one ($\alpha_{full}$) expected for full polarization at each ν. At a fixed magnetic field $B$, $\alpha_{full}$ is proportional to the density $n_P$ of electrons that can contribute to the spin polarization. As the cyclotron energy separating the $N = 0$ and $N = 1$ Landau levels is much larger than the Zeeman energy in GaAs, there should be no occupied states in the $N = 1$ Landau level at ν < 2 and no unoccupied states in the $N = 0$ Landau level at ν > 2. With this constraint, $n_P$ and hence $\alpha_{full}$ are given in the form of

$$\alpha_{full}(\nu) = c \cdot n_P(\nu) = \begin{cases} c \cdot \dfrac{eB}{h} \cdot \nu \ (= c \cdot n) & \text{(for } 0 \leq \nu \leq 1) \\ c \cdot \dfrac{eB}{h} \cdot |\nu - 2| & \text{(for } 1 < \nu \leq 3), \end{cases} \quad (2)$$

where $eB/h$ is the degeneracy of a spin-split Landau level. The constant $c$ in Eq. (2) can be determined using the value of $\alpha_\nu$ that is experimentally obtained for a fully polarized state. Here, we use the ν = 5/3 as a reference, i.e., $P = \alpha_{5/3}/\alpha_{full}(5/3) = 1$. The function $\alpha_{full}(\nu)$ thus obtained is shown as a dashed line in Fig. 3A. Equation (2) also accounts for the linear $B$ dependence of $\alpha$ at ν = 5/2 shown in Fig. 2A.

**Estimation of error bars**

The main cause of the error in our measurements is the uncertainty of the reference frequency $f_0$, set by the stability of the magnetic field. The error bars in the plot of $\alpha$ in Figs. 2-4 represent the variance caused by a fluctuation in the magnetic field by ±0.1 mT, which produces an uncertainty of the reference frequency by ± 0.7 kHz at 6.4 T. Although the measured RD-NMR spectra contain some data scattering, we find that the error due to this scattering is minor as compared to the error due to the magnetic field. We confirmed this by repeating measurements and observing that the obtained values of $\alpha$ are reproducible. In some experimental conditions, the RD-NMR signal becomes low, for example at large ν close to 8/3, as a result of the large Knight shift and the small overlap between $|\psi_\nu(z)|^2$ and $|\psi_{read}(z)|^2$ (See Fig. S1A and S1B). In such cases, we averaged the spectra of multiple repeated measurements so that the relevant error becomes sufficiently small.

The spin polarization $P$, on the other hand, is calculated as the ratio between the obtained value of $\alpha$ and the one expected for the maximum polarization. Thus, the error in $P$ becomes large (small) for those states around ν = 2 (ν = 1), for which the maximum polarization is small (large). Although the absolute value of $\alpha$ is subject to the uncertainty (± 5%) of the well width, the spin polarization, which relies only on the relative values of $\alpha$ for different ν, is barely affected.

**Polarization at ν = 1 and 1/3**

In Fig. 3A, the values of $\alpha$ at ν = 1 and 1/3 are found to be 88-89% of the expected maximal values estimated based on $\alpha$ at ν = 5/3. We believe that this is due to the limitations in applying the same fitting procedure to such a diverse range of filling factors (gate voltages). As seen in Fig. S1A, at 6.4 T the electron density distribution $|\psi_\nu(z)|^2$ becomes nearly symmetric at ν = 2, whereas at ν = 1 and 1/3 it is strongly squeezed, with much larger overlap with $|\psi_{read}(z)|^2$ as compared to the cases for larger ν. As seen in Fig. S2, the Knight shift observed at ν = 1/3 is actually larger than that at ν = 5/3, consistent with the model (Fig. S1B). The deviation of $\alpha$ at ν = 1/3 from the expected value therefore suggests that the model overcompensates the effect of $|\psi_\nu(z)|^2$. A recent optical experiment reports that full polarization at ν = 1 occurs only within a very narrow range of filling factor (*SR5*), which may partly account for our observation at ν = 1. However, we emphasize that using $\alpha$ at ν = 1 and 1/3 as a reference for full polarization instead would lead to the unphysical results that the polarization exceeds 100% at ν = 5/3 and in the $N = 1$ Landau level.

**Analysis based on non-interacting composite-fermion model**

According to the composite-fermion (CF) model, at half fillings CFs experience zero magnetic field in the mean-field level, forming a compressible liquid state with a Fermi surface. In the non-interacting CF model, the energy dispersion of CFs can be described by a constant effective mass $m_{CF}$. When the system is fully polarized at sufficiently high fields, the Fermi energy of CFs is then given by $\varepsilon_{CF}^* = n_{CF}/D_{CF}$, with $D_{CF} = m_{CF}/(2\pi\hbar^2)$ the density of states and $n_{CF}$ the density of CFs ($n_{CF} = n, n/3, n/5$ for ν = 1/2, 3/2, and 5/2, respectively, with $n$ the density of electrons). When the Zeeman energy $E_Z$ becomes smaller than $\varepsilon_{CF}^*$ at low fields, both spin-up (↑) and spin-down (↓) bands are occupied up to the Fermi level with their band bottoms offset by $E_Z$. Consequently, the spin-up and spin-down bands have their respective Fermi energies given by $\varepsilon_{CF}^{\uparrow(\downarrow)} = n_{CF}^{\uparrow(\downarrow)}/D_{CF}$, with $\varepsilon_{CF}^\uparrow - \varepsilon_{CF}^\downarrow = E_Z$ and $\varepsilon_{CF}^\uparrow + \varepsilon_{CF}^\downarrow = \varepsilon_{CF}^*$, where $n_{CF}^{\uparrow(\downarrow)}$ is the density of spin-up (spin-down) CFs. At $T = 0$, the spin polarization $P = (n_{CF}^\uparrow - n_{CF}^\downarrow)/(n_{CF}^\uparrow +$

$n_{CF}^\downarrow$) is therefore given by a simple formula $P = E_Z/\varepsilon_{CF}^*$ for $E_Z \leq \varepsilon_{CF}^*$. The spin polarizations $P = 0.53$ and $0.28$ at 6.4 T we measured for $\nu = 1/2$ and $3/2$ at the base temperature correspond to $m_{CF} = 1.24 m_e$ and $0.65 m_e$, respectively, while the full polarization of $\nu = 5/2$ implies $m_{CF} \geq 2.30 m_e$.

With the non-interacting CF model, the Fermi-Dirac distribution of CFs leads to the temperature dependence of the spin polarization given in the form of (*SR6-SR8*)

$$P(T) = \frac{1}{\varepsilon_{CF}^*}\left[E_Z - 2k_B T \sinh^{-1}\left(\frac{\sinh(E_Z/2k_B T)}{\exp(\varepsilon_{CF}^*/2k_B T)}\right)\right], \quad (3)$$

where $k_B$ is the Boltzmann constant. Fitting the temperature-dependent data at 6.4 T in Fig. 4C using the above equation yields $m_{CF} = (2.68 \pm 0.24)m_e$ and $(0.71 \pm 0.05)m_e$ for $\nu = 5/2$ and $3/2$, respectively.

With the assumption that $m_{CF}$ scales as $m_{CF}/m_e = a\sqrt{B}$ ($a$ is constant and $B$ in Tesla) (*SR9*), our results can be compared with the values in the literature obtained at different magnetic fields. Our results at 6.4 T imply $m_{CF}/m_e = 0.49\sqrt{B}$, $(0.28 \pm 0.02)\sqrt{B}$, $(1.06 \pm 0.09)\sqrt{B}$ for $\nu = 1/2$, 3/2, and 5/2, respectively. We note that our value for $\nu = 1/2$ is consistent with, but slightly smaller than, the previously reported values (*SR7, SR10, SR11*). This is because in our experiment $\nu = 1/2$ is accessed by applying a large negative voltage to the *back* gate, which squeezes the wave function and enhances the Coulomb interaction, making $m_{CF}$ smaller (*SR12*). This contrasts with the case in Ref. (*SR11*), where a *front* gate was used to tune the electron density, which results in the opposite relation between the width of the 2DES and the electron density.

On the other hand, the difference in $m_{CF}$ measured for $\nu = 1/2$ and $3/2$ in our experiment, also noted in Ref. (*SR10*), is not explained by the wave function squeezing. As also pointed out in Ref. (*SR10*), the smaller $m_{CF}$ for $\nu = 3/2$ than for $\nu = 1/2$ is consistent with the much larger magnetic field to fully polarize $\nu = 4/3$ than $\nu = 2/3$.

Fermi surface properties at $\nu = 5/2$ have been reported in surface acoustic measurements at temperatures where the $\nu = 5/2$ FQH state has vanished (*SR13*). The measurements, however, suggest that $\nu = 5/2$ and $3/2$ are similar in spin polarization. Such discrepancy between geometrical resonance experiments and direct spin polarization measurements has been known and remains to be resolved.

**Figures**

Fig. S1
**A**. Numerically calculated probability density distribution of subband wave functions at each $\nu$ (red solid lines) and at $\nu_{read} = 0.59$ (black dotted lines) ($B = 6.4$ T). **B**. Spectra at $B = 6.4$ T calculated for various $\nu$ with $\nu_{read} = 0.59$ and $\Gamma = 1.5$ kHz. **(Top panel)** Spectra for $\nu = 5/2$ calculated using different values of $\alpha$ corresponding to a polarization between 0 and 1. **(Middle panel)** Spectra for $\nu = 1/3$, 5/3, and 7/3 calculated with the same value of $\alpha$ corresponding to full polarization of these states. All parameters except $\psi_\nu(z)$ are the same. **(Bottom panel)** Spectra for $\nu = 1$ and 8/3 calculated for the case of full polarization.

Fig. S2
RD-NMR spectra at various filling factors ranging from $\nu = 1/3$ to 8/3, measured at 6.4 T and 12 mK. The black solid lines show the results of fitting.

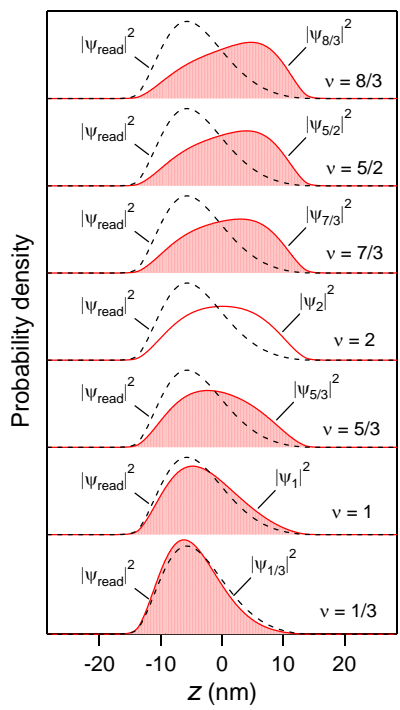 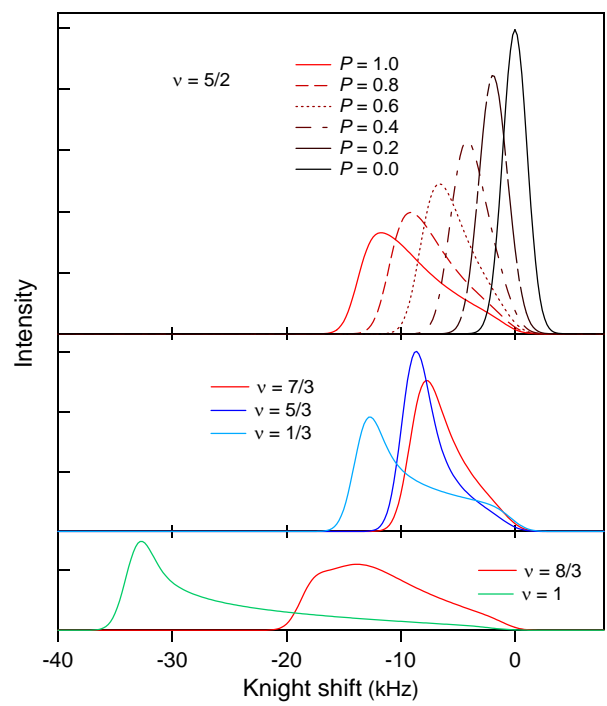

Fig. S1

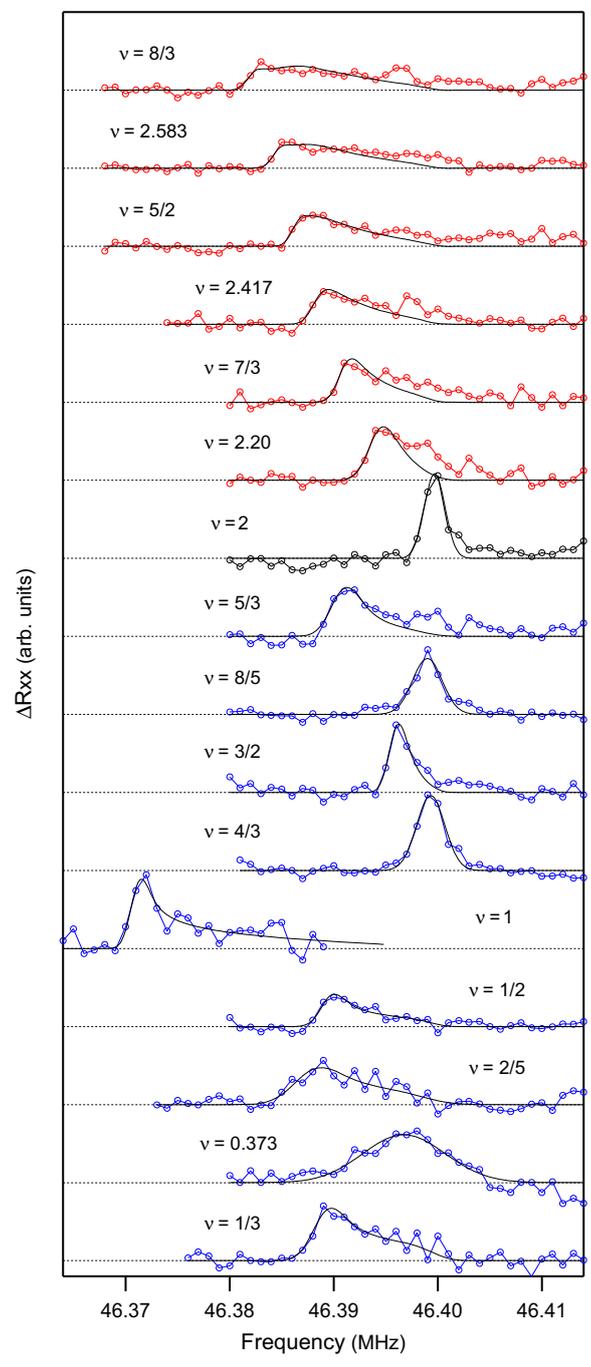

Fig. S2